# Electron Scattering in Liquid Water and Amorphous Ice: A Striking Resemblance

Ruth Signorell

Department of Chemistry and Applied Biosciences, ETH Zürich, Vladimir-Prelog-Weg 2. CH-8093 Zürich, Switzerland

*Correspondence to: rsignorell@ethz.ch.

**Abstract**: The lack of accurate low-energy electron scattering cross sections for liquid water is a substantial source of uncertainty in the modelling of radiation chemistry and biology. The use of existing amorphous ice scattering cross sections for lack of liquid data has been discussed controversially over decades. Here, we compare experimental photoemission data of liquid water with corresponding predictions using amorphous ice cross sections, with the aim of resolving the debate regarding the difference of electron scattering in liquid water and amorphous ice. We find very similar scattering properties in the liquid and the ice for electron kinetic energies up to a few hundred electron volts. The scattering cross sections recommended here for liquid water are an extension of the amorphous ice cross sections. Within the framework of currently available experimental data, our work answers one of the most debated questions regarding electron scattering in liquid water.

Low-energy electron scattering in liquid water is recognized as having important implications in a wide range of fields, including radiation damage in aqueous systems [1-9]. These electrons are abundantly produced during slowing-down processes from various precursor processes. The quantitative description of the transport properties of sub-keV electrons in liquid water, however, has been very challenging due to the lack of reliable electron scattering cross sections (CSs) for the liquid and related quantities, such as electron inelastic and elastic mean free paths (referred to as IMFPs and EMPFs, respectively).

Various models have been proposed for the prediction of mean free paths (MFPs), many of which are based on different treatments of the dielectric response (see e.g. [5,10-16] and references therein). It has been shown that predicted MFPs are very sensitive to the particular model used for electron kinetic energies (eKEs) below a few hundred eV, with the biggest deviations in the range below a few 10eV where the appropriateness of some model assumptions are disputed. The situation in the subexcitation range (eKEs ≤ 7eV) remains particularly uncertain. Model approaches are largely missing in this range due to the complexity of the relevant processes to be described.

The retrieval of CSs from experimental data has been severely hampered by difficulties in performing experiments with low energy electrons for liquid water. Liquid bulk water is not compatible with the vacuum required for such experiments. Substantial progress has been made in recent years through the invention of liquid water microjet photoelectron spectroscopy [17] and droplet photoelectron imaging [18]. This has resulted in experimental electron attenuation lengths (EALs) [19-21], photoelectron anisotropy parameters [22-24] and liquid CSs in the subexcitation range determined from droplet photoelectron images [18], but not yet in detailed CSs for the liquid that cover the whole sub-keV range. Detailed CSs, i. e. multiple differential CSs and energetics, exist only for amorphous ice retrieved from experimental ice data by Sanche and coworkers for the entire range below 100eV eKE [25,26]. The use of these ice CSs for lack of liquid CSs has become one of the most controversially discussed issues in the field; so far with no clear outcome. This controversy mainly resulted from the fact that some experimental observations in the liquid that are indirectly related to scattering CSs could seemingly not be modelled accurately enough with the ice CSs. Differences between theoretically modelled liquid MFPs and those derived from the experimental amorphous ice CSs also contributed to the dispute. Ad hoc scaling factors for CSs have been proposed to account for these potential differences between ice and liquid (see e.g. [5,6,12,27,28] and references therein). Even though firm physical arguments in favor of rather than against a close resemblance of liquid and ice CSs have been put forward (e.g. refs. [25,26]), a confirmation based on a quantitative assessment of experimental data has not been attempted so far. In this work, we combine the most reliable experimental information available from photoelectron spectroscopy of liquid water microjets and water droplets with detailed electron scattering simulations using ice CSs to resolve this issue. The goal is to provide a recommendation for electron scattering CSs for liquid water in the entire sub-keV range.

Experimentally determined effective electron attenuation lengths $EAL^{eff}$ [19], anisotropy parameters β [22-24] and photoelectron velocity map images (VMIs) [18] from liquid water studies are compared with corresponding predictions using the amorphous ice CSs from refs. [25,26]. The latter were retrieved from electron energy loss spectra recorded for amorphous ice films. The EAL$^{eff}$ and β parameters were obtained from liquid water micro-jet measurements, while the VMIs were recorded for water droplets. These liquid measurements were performed at liquid temperature in the range of ~240-270K (i. e. for supercooled water). The calculated EAL$^{eff}$, β parameters and VMIs are obtained from a detailed electron scattering model based on a Monte-Carlo solution of the transport equation using extended ice CSs (refs. [18,24-26,29] and ref. [30] sections S1-S5). The model reproduces the exact experimental conditions of each respective experiment; i. e. relevant experimental arrangements, intensity distribution of the ionizing radiation in the liquid samples, and electron sampling conditions.

Fig. 1 compares the experimental effective attenuation length for liquid water from ref. [19], EAL$_S^{eff}$, with a Monte Carlo prediction, EAL$_M^{eff}$, using the electron scattering CSs of amorphous ice from refs. [25,26] for eKEs below 100 eV. The definition of EAL$^{eff}$ deviates from the usual definition of an EAL (ref. [30] section S1). In essence, the pioneering experiments of ref. [19] amounted to an elegant measurement of the absolute photoemission yield for liquid water, which was converted into an effective EAL$_S^{eff}$. In the eKE range covered by the amorphous ice data (eKE < 100 eV) EAL$_S^{eff}$ and EAL$_M^{eff}$ agree very well within their respective uncertainties – and, there is no systematic deviation between the two, e.g. one of them consistently higher than the other. Evidently, electron scattering is quite similar in water and ice, and the amorphous ice CSs from refs. [25,26] are clearly adequate to predict properties of the liquid. Even though the authors of ref. [19] emphasize that background instabilities make their results less reliable below eKEs ~10eV, we show these data in Fig. 1 for completeness. Secondary electrons, phonon scattering and background effects are major issues in this energy region (ref. [30] section S1).

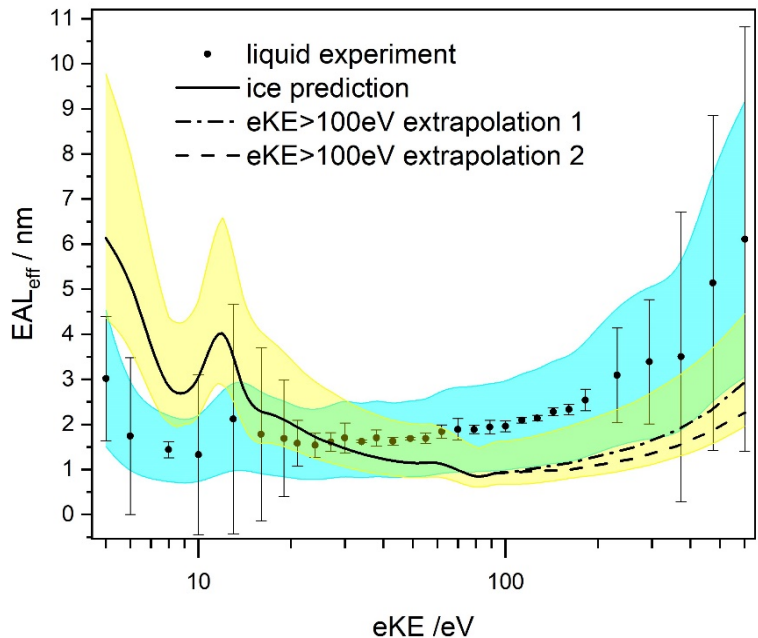

**Fig. 1. Effective attenuation length EAL$^{eff}$.** Experimental effective attenuation length EAL$_S^{eff}$ for liquid water from ref. [19] (black dots). The black error bars for EAL$_S^{eff}$ indicate three standard deviations of data resulting from measurements on different days [19], while the blue shaded area represents an estimate of additional systematic uncertainties arising from various sources (ref. [30] section S1). Both contributions determine the total uncertainty of EAL$_S^{eff}$.

Monte Carlo prediction of the effective attenuation lenght $EAL_M^{eff}$ using electron scattering CSs derived from experiments on amorphous ice [25,26] (full black line). The CSs above 100eV eKE (dashed-dotted line) are obtained by extrapolation with the model from ref. [10]. The uncertainty for $EAL_M^{eff}$ (shaded yellow area) corresponds to an uncertainty of the absolute total CSs of 45% [25,26]. The dashed line is obtained by extrapolation of the CSs with the model from ref. [11].

Experimental amorphous ice CSs are only available up to 100 eV [25]. For calculations beyond that range, we therefore have to follow a different line in order to compare with the $EAL_S^{eff}$ of ref. [19] up to ~600eV. Given the large total uncertainties of the $EAL_S^{eff}$ data for eKE > 100eV (Fig. 1), nothing would be gained by extracting liquid CSs from a fit to the $EAL_S^{eff}$. Instead, we suggest the approach described in ref. [30] section S5 to extrapolate liquid CSs for eKE above 100eV in order to predict $EAL^{eff}$ in that range. The approach uses the model for liquid water of ref. [10] – taking only the energy dependence of the electronically inelastic mean free path, rather than its absolute values - to extrapolate the differential CSs for amorphous ice at 100eV from ref. [25] to higher eKE values. Fig. 1 compares the $EAL_M^{eff}$ (dashed-dotted line) thus predicted with the $EAL_S^{eff}$ from ref. [19]. Within uncertainties, the extrapolation yields a similarly good agreement between the two $EAL^{eff}$ data sets above 100 eV, as was obtained with the amorphous ice data for eKEs below 100eV. The model of ref. [10] is based on calculations of the IMFP from the optical energy-loss function (ELF) using the relativistic full Penn algorithm. The theoretically predicted absolute values were scaled by factor of ~3.1 to match the (lower) electron loss CSs at 100eV of ref. [25]. This scaling factor reduces to ~1.8 if the model from ref. [11] (dashed line, Fig. 1) is used instead of ref. [10]. The reduction originates from the inclusion of exchange and correlation effects in the former model in ref. [11].

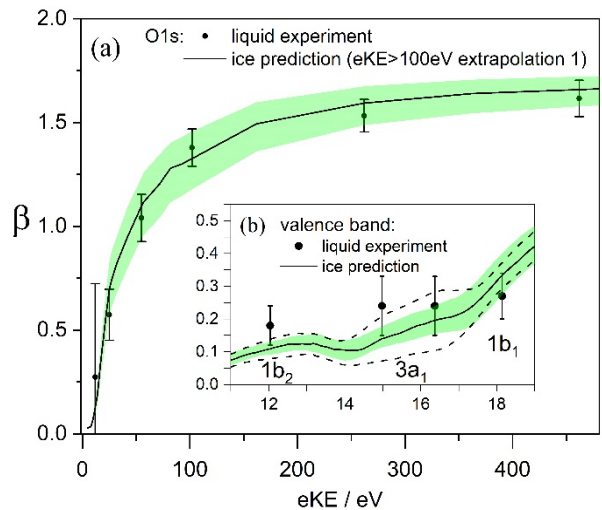

**Fig. 2. Anisotropy parameters β. (A)** Experimental $β_T$ for ionization from the O1s orbital of liquid water from ref. [22] (black dots). Monte Carlo prediction $β_M$ for ionization from O1s using electron scattering cross sections derived from experiments on amorphous ice [25,26] (full black line). The black error bars for $β_T$ indicate three standard deviations quoted in ref. [22]. The uncertainty for $β_M$ (shaded green area) corresponds to an uncertainty of 20% in the relative cross sections for electronically inelastic and quasi-elastic scattering [25,26]. **(B)** Experimental $\beta_N$ for ionization from the three valence orbitals $1b_1$, $3a_1$ and $1b_2$ of liquid water from ref. [23] (black dots). Monte Carlo prediction $\beta_M$ for ionization from the valence orbitals using electron scattering cross sections derived from experiments on amorphous ice [25,26] and water cluster data [24] (full black line). The black error bars for $\beta_N$ indicate the statistical errors quoted in ref. [23]. The uncertainty for $\beta_M$ (shaded green area) is the same as in A. The dashed lines represent uncertainties arising from the uncertainty of the hexamer water cluster [24].

The photoelectron anisotropy parameters $\beta_T$ and $\beta_N$ recorded for liquid water in refs. [22,23] provide a second experimental data set for the comparison of electron scattering in liquid water and amorphous ice for eKEs above 10eV. The results are illustrated in Fig. 2A for ionization from the O1s orbital and in Fig. 2B for ionization from the valence orbitals. The predictions $\beta_M$ based on the amorphous ice CSs from refs. [25,26] are described in ref. [30] section S2. For the calculated $\beta_M$ above eKEs of 100eV, we use the same extrapolation as described above for $EAL_M^{eff}$ with the model from ref. [10]. The very good agreement between liquid water and ice β-parameters (within the uncertainties indicated in Fig. 2) confirms the similarity of electron scattering in the two phases already found in Fig. 1 for $EAL^{eff}$. Again, no meaningful improvement of the CSs could be gained from fitting to the experimental β-parameters.

Reliable information on sub-excitation electron scattering became recently available from VMI photoelectron studies of droplets after EUV excitation at more than 40 different photon energies below ~15.4eV [18,29]. The droplet VMIs (Fig. 3A) contain information on the eKE and entire photoelectron angular distribution (PAD), which goes beyond that of a single β-parameter. The cross sections for liquid water were directly extracted from fits of calculated droplet VMIs to experimental VMIs as described in refs. [18,29]. It turned out that the experimental droplet VMIs were best reproduced by the simulations when fixing the liquid CSs at the supporting points to those of amorphous ice [25] without further refinement, indicating that the liquid water and the amorphous ice cross sections are very similar in the sub-excitation regime. As explained in ref. [30] sections S3 and S5, we have slightly refined the previous liquid CSs [18,29] using the entire information for ice in refs. [25,26]. We determine somewhat higher uncertainties for the absolute values of the liquid CSs (~factor of two) compared with ice CSs in the sub-excitation regime (30%). The close agreement between liquid and ice is visualized in Fig. 3B for the total MFP (TMFP); i. e. the combined MFP of all inelastic and isotropic elastic (momentum transfer CS, transport CS) contributions. The liquid water TMFPs (full black line) have larger uncertainties (shaded green area) compared with amorphous ice (black dots with error bars). The total IMPF (dashed-dotted line) and the isotropic EMPF (dashed line) of liquid water are also indicated in Fig. 3B. Further confirmation of the close similarity between liquid and ice CSs is provided by the good agreement between thermalization lengths measured in the liquid [31] with predictions using the amorphous ice CSs [12,28] for eKEs below ~4eV. Ad hoc upscaling ice CSs by a factor of two, as suggested in refs. [12,27,28] to better represent liquid CSs, in fact deteriorated the agreement with the experimental thermalization lengths in ref. [12].

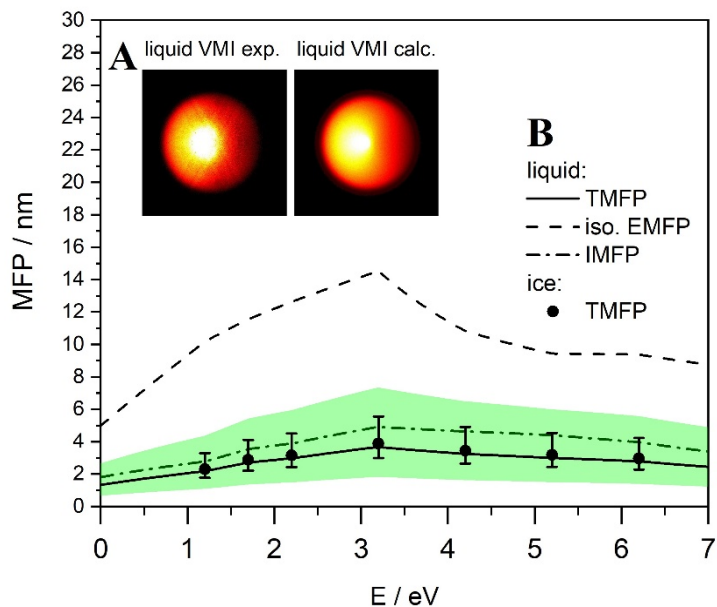

**Fig. 3**. **Mean free paths (MFPs) for subexcitation electrons. (A)** VMI photoelectron spectra of water droplets recorded at a photon energy of 14.9eV (see also [18,32]). Left: experimental spectrum. Right: spectrum simulated with liquid CSs (ref. [30] section S5). **(B)** TMFP of amorphous ice (black dots) with uncertainties (black error bars) from refs. [25,26] and of liquid water (black full line) with uncertainties (green shaded area) (ref. [30] section S3). Total IMFP (black dashed-dotted line) and isotropic EMFP (black dashed line) of liquid water. The total IMPF includes all inelastic channels (electronic, vibrational, phonon). E is the electron's total energy with respect to the vacuum level [25,26].

Currently known observables ($EAL^{eff}$, β, VMI) characteristic of the scattering behavior of electrons in liquid water below eKEs of 100eV can be predicted with CSs for amorphous ice within the quoted uncertainties (Figs. 1-3). Above 100eV, the scattering behavior of electrons is similarly well predicted by the suggested extrapolation of the amorphous ice CSs on the basis of the functional form of the energy dependence predicted theoretically for liquid water [10]. All of these results taken together lead to the expectation that the electron scattering cross sections in liquid water and ice agree at least within a factor of two. Currently available experimental data for the liquid are not sufficiently accurate to bracket this even more narrowly. Fitting the CSs for the liquid to current experimental data (Figs.1 and 2) would not offer any improvement over using the recommended liquid CSs described in ref. [30] section S5. The agreement between experimental liquid and ice data reported here is generally closer than the one between experimental and theoretical liquid data. Based on the following general physical considerations, already outlined in refs. [25,26], one might in fact expect an even closer agreement between liquid and ice CSs than current uncertainties suggest. The major differences between electron scattering in the gas phase (isolated molecules) and in the condensed phase arise from the change in the density of the scattering centers and from the intermolecular interactions. Short range interactions (mainly hydrogen bonds) give rise to low energy phonons (hindered rotations and translations), while long-range effects - mainly through dielectric screening – directly correlate with the density. Intermolecular interactions are relatively weak (on the order of 0.1eV or less) and mostly affect the phonon spectrum (and thus phonon scattering). These interactions are too weak (percent of electronic energies) to affect the electronic structure significantly. Apart from these energetic effects (the interaction strength with a scattering center) electron scattering in the condensed phase is also influenced by the interference between neighbouring scattering events, a direct consequence of the increased density. It is important to note that such interferences are naturally contained in experimental CSs and hence also in the effective CSs reported here (see also [25,26]). As the density and the degree of disorder are very similar in liquid water and amorphous ice, the same should hold for interaction strengths, screening and interference behavior. Consequently, one expects very similar CSs in the liquid and the amorphous ice. The only remaining effect that could cause a difference between scattering in water and ice is the difference in temperature. The effect would potentially be most pronounced in the subexcitation regime, where phonons are important. At higher temperatures, thermally populated excited phonon levels could cause an increase of eKE in a scattering event. On average this could reduce the energy loss per scattering event for the low energy phonon channels in the liquid. At the thermal energies of the liquids considered here (on the order of 0.01eV), this effect is not expected to be very pronounced. This is also in agreement with the fact that there is actually no evidence for larger differences between ice and liquid in the subexcitation range (Fig. 3). Finally, our results exclude proposed ad hoc scaling factors of more than two between ice and liquid CSs, both for inelastic and isotropic (transport) elastic CSs, and they also provide evidence against the use of smaller scaling factors (ref. [30] section S5).

The present work resolves the controversy regarding the difference in scattering behavior of electrons in liquid water and amorphous ice, based on experimental data for liquid and ice. The close similarity of electron scattering in the liquid and in ice finally allows one to bracket the range of scattering cross sections for the liquid. Compelling evidence is provided that the previously available ice cross sections from Sanche and coworkers [25,26] with the extensions described in the present work currently provide the most reliable cross sections for the liquid (ref. [30], section S5, Table S1), with maximum uncertainties on the order of a factor of two. Ad hoc scaling of the ice cross sections as previously suggested is clearly not recommended. The present results are expected to have far-reaching implications for the modelling of electron scattering in aqueous environments, and thus for the understanding of chemical and cellular radiation damage.

**Acknowledgments:** This project has received funding from the European Union's Horizon 2020 research and innovation program from the European Research Council under the Grant Agreement No 786636, and the research was supported by the NCCR MUST, funded by the Swiss National Science Foundation (SNSF), through ETH-FAST, and through SNSF project no. 200020_172472. The author thanks all coworkers and collaborators who have contributed to the previous experimental and theoretical work that is discussed in the current publication. The author is also a grateful recipient of a Humboldt Research Prize form the Alexander von Humboldt Foundation and a Mildred Dresselhaus Guestprofessorship from the Centre for Ultrafast Imaging in Hamburg.

# Supplemental Material

## Electron Scattering in Liquid Water and Amorphous Ice: A Striking Resemblance

Ruth Signorell

correspondence to: rsignorell@ethz.ch

# S1. Definition, error estimate, modelling of $EAL^{eff}$

**Definition of the experimental $EAL_S^{eff}$:**

The effective $EAL_S^{eff}$ as defined in ref. [19] corresponds to an average effective probing depth, and does not follow the usual definition of the EAL. The difference between the two definitions amounts to a factor on the order of two (see below). In ref. [19], the effective $EAL_S^{eff}$ is not measured directly. The actual measurement quantity is the photoelectron yield inferred from the relative intensities of the O1s photoemission bands from a liquid jet and its surrounding gas phase, respectively. On the basis of simplifying assumptions about the photoemission process, a simple geometric model is then used to convert the photoelectron yield into an effective EAL.

Following the notation of ref. [19], the observed band intensities are given by

$$I_{g/l}(hv) = c_{g/l}(eKE) n_{g/l}(hv) \sigma_{g/l}(hv) f(hv)$$

The subscripts $g$ and $l$ refer to the gas and liquid phase, respectively. $eKE = hv - eBE_{g/l}$ is the photoelectron kinetic energy, $hv$ is the photon energy, and $eBE_{g/l}$ is the electron binding energy. $I_{g/l}$ is the observed photoelectron intensity (integrated over the O1s emission band), $c_{g/l}$ is a sensitivity factor accounting for the anisotropy of the photoemission, $n_{g/l}$ is the number of ionization events, that give rise to detectable photoemission, $\sigma_{g/l}$ is the photoionization cross section, and $f$ is the incoming photon flux. The number of photoelectrons emitted is given by the distributions of molecular density $\overline{\rho_{g/l}}$ and photon flux $\overline{f_{g/l}}$ in the detectable volume and by the photoemission yield $\gamma_{g/l}$

$$n_{g/l}(hv) = \rho_{g/l} \cdot \langle \overline{\rho_{g/l}}(x,y,z) \cdot \overline{f_{g/l}}(hv,x,y,z) \cdot \gamma_{g/l}(eKE,x,y,z) \rangle_{g/l}$$

The brackets indicate integration over the detectable volume in the gas and in the liquid, respectively. The bars indicate normalization to the saturation density $\rho_{g,l}$ and incoming photon flux $f(hv)$, respectively.

In the gas, the photon flux distribution is uniform, so that $n_g$ is only determined by the vapor density distribution and the irradiated gas volume (see Eq.4 in ref. [19]). The yield is assumed to be approximately unity, $\gamma_g \approx 1$ (neglecting scattering and reabsorption of electrons in the gas phase).

$$n_g(hv) \approx \rho_g \cdot \langle \overline{\rho_g}(x,y,z) \rangle_g$$

In the liquid, by contrast, the molecular density is uniform, while the intensity distribution is not. Moreover, the yield is energy dependent and less then unity (because of scattering).

$$n_l(hv) = \rho_l \cdot \langle \overline{f_l}(hv,x,y,z) \cdot \gamma_l(eKE,x,y,z) \rangle_l$$

In ref. [19], the integration over the detectable volume is replaced by the corresponding integration in the (x,y)-plane perpendicular to the liquid jet (z) at the focal position of the X-ray beam. This is indicated in the following by dropping the z-coordinate. $\langle \overline{f_l}(hv,x,y) \cdot \gamma_l(eKE,x,y) \rangle_l$ is now equated to a quarter section of the surface layer of the liquid jet whose thickness is considered an « effective EAL » (see Fig.3b of ref. [19])

$$\langle \overline{f_l}(h\nu, x, y) \cdot \gamma_l(E, x, y) \rangle_l = \frac{\pi}{2} r_0 \cdot EAL_S^{eff}$$

where $r_0$ is the liquid jet radius. This effective EAL is typically significantly smaller than either the usual EAL or the average probing depth (the latter often being taken as an approximate measure of the EAL), because the overall photoemission yield is less than unity even for ionization very close to the surface (half of the photoelectrons formed by ionization initially move towards the inside of the liquid).

The experimental value for the effective EAL quoted in ref. [19] is given by

$$EAL_S^{eff} = \frac{1}{A(eKE)} \frac{I_l(eKE)}{I_g(eKE)}$$

with the « instrumental correction factor »

$$A(eKE) = \frac{\pi}{2} r_0 \frac{c_l(eKE)\sigma_l(h\nu)\rho_l(eKE)}{c_g(eKE)\sigma_g(h\nu)n_g(h\nu)}$$

In the experiment, the partially overlapping gas and liquid O1s bands appear on a continuous background of inelastically scattered valence photoelectrons and secondary electrons. This background was subtracted before retrieving the relative intensities $I_{g/l}$ from appropriate band shape fits, so that simulations of the experimental results need not account for scattered valence electrons or for secondary electrons. However, the authors of ref. [19] note the increasing uncertainty of the background subtraction at low eKEs ($\lesssim 10$ eV). A large part of this uncertainty arises from the increased phonon and vibron scattering in the liquid smearing out the band shape. As a result, a significant part of quasielastically scattered photoelectrons becomes indistinguishable from the background of scattered valence and secondary electrons. This in turn reduces the observed yield $\gamma_l$ of photoemission from the liquid and thus $EAL_S^{eff}$.

**Main sources of experimental error:**

We considered the following error sources to estimate the overall error of the experimental values for the effective $EAL_S^{eff}$.

- Saturated water vapor density $\rho_g$ : -28%/+38%, quoted in ref. [19]

- Vapor density profile $\overline{\rho_g}$ $(x, y, z)$: 5-10%, estimated from figure S4 in ref. [19]

- X-ray beam profile: 10% for radius, estimated from figure S1 in ref. [19], plus an estimated 10% for the radial intensity profile

- Liquid jet radius $r_0$ : 10-20%, estimated from figure S2 in ref. [19]

- Absorption cross section ratio $\sigma_l/\sigma_g$: estimated 5-10%

- Sensitivity factors $c_{g/l}$ : 10% each, estimated from uncertainties of $\beta$ values

- O1s band intensity ratio $I_l/I_g$ : 5-10%, estimated from figure 2 of ref. [19] (not including the additional uncertainty at eKE$\lesssim$ 10 eV mentioned above), plus 13% quoted in ref. [19] for the uncertainty due to scattering of photoelectrons in the vapor

We estimate that these errors add up to an additional systematic uncertainty for the $\text{EAL}_\text{S}^\text{eff}$ of at least a factor of two. The *total* uncertainty corresponds to this additional uncertainty (blue shaded area in Fig. 1 in the main manuscript) on top of the standard deviation of data resulting from measurements on different days (black error bars in Fig. 1 in the main manuscript).

**Modelling of $\text{EAL}_\text{M}^\text{eff}$:**

The model mimics the experiment with the calculation of $\text{EAL}_\text{M}^\text{eff}$ following the definition of $\text{EAL}_\text{S}^\text{eff}$. The light intensity distribution inside the liquid jet is calculated from the complex index of refraction. Since the real part is very close to unity over the whole range of photon energies considered (550-1140 eV) the flux distribution is simply given by the exponential damping in the direction of propagation starting from the point of incidence. The damping length is given by $\lambda(h\nu)/4\pi k(h\nu)$, where $\lambda(h\nu)$ is the vacuum wavelength and $k(h\nu)$ the imaginary part of the complex index of refraction. The detectable volume is the half of the jet facing the detector. The $\text{EAL}_\text{M}^\text{eff}$ is thus given in terms of the overall photoemission yield $\gamma_{l,tot}$ of the simulation (linear polarization of the X-ray beam parallel to the axis of detection).

$$\text{EAL}_\text{M}^\text{eff} = \frac{2}{\pi r_0} \cdot \gamma_{l,tot} \cdot \langle \overline{f_l}(h\nu, x, y) \rangle_l$$

with

$$\gamma_{l,tot} = \langle \overline{f_l}(h\nu, x, y) \cdot \gamma_l(E, x, y) \rangle_l / \langle \overline{f_l}(h\nu, x, y) \rangle_l$$

The Monte Carlo model for electron scattering described in section S4 and the recommended liquid CSs described in section S5 are used for the scattering calculations. To account for the effect of phonon and vibron scattering on the apparent photoemission yield from the liquid actually observed in the experiment (see above), we decompose the photoemission band shape predicted by our model into a leading Gaussian and a (broader) background as illustrated in Fig. S1. The leading Gaussian is generally well defined by the high eKE band edge even in the eKE-range below 10 eV. Only the leading Gaussian is assumed to contribute to the observed band intensity $I_l$ and thus to $\gamma_{l,tot}$. Note that the total photoemission signal shown in Fig. S1 appears in the experiment on top of a large background of scattered valence band electrons and secondary electrons (through electron impact ionization) rising steeply towards small eKE. Therefore, this is a critical region and experimental results are rather sensitive to the treatment of the background.

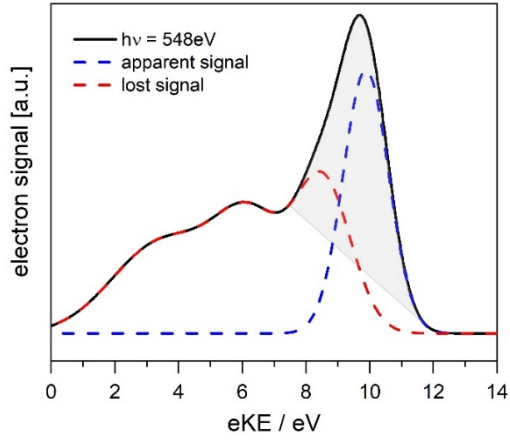

**Fig. S1. The effect of phonon/vibron scattering on the apparent photoemission signal.**
Simulated photoemission signal from the liquid jet at hν = 548 eV. Decomposition of the full signal into apparent signal (leading Gaussian, blue dashed) and the part of the signal lost in the background (red dashed). The gray shaded area illustrates the part of the total signal effectively contributing to the photoemission yield from the liquid, $\gamma_{l,tot}$.

## S2. Definition, experiment and modelling of $\beta$ parameter

$\beta$ describes the angular distribution of the photoemission from a sample irradiated by linearly polarized light,

$$I(\theta) \propto 1 + \beta P_2(\cos\theta).$$

$P_2$ is the 2$^{nd}$ Legendre polynomial and $\theta$ is the angle between the directions of polarization and detection.

### Experimental $\beta_T$ and $\beta_N$:

The determination of $\beta_T$ and $\beta_N$ from the experimental photoelectron spectra recorded at different linear polarization directions of the ionizing light w.r.t. to the axis of detection is described in refs. [22] and [23], respectively. All experimental data are retrieved from liquid water micro-jet studies [22,23].

### Modelling of $\beta_M$:

The model mimics the experiment and the calculation of $\beta_M$ follows the definition of $\beta_T$ for O1s ionization and $\beta_N$ for valence ionization. The light intensity distribution inside the liquid jet is calculated from the complex index of refraction. $\beta_M$ is calculated from the simulated photoelectron spectra with linear polarization at 0 and 90 degree, respectively, w.r.t. the axis of detection (90 degree means parallel to the liquid jet axis). The opening angle for detection was assumed to be 16 degrees. The Monte Carlo model described in section S4 and the recommended liquid CSs described in section S5 are used for the scattering calculations. For the initial angular distribution of photoelectrons in the liquid (the „genuine" distribution) for ionization from the O1s orbital, we used the experimental gas phase anisotropy parameters from ref. [22], with double logarithmic interpolation between experimental data points. The genuine binding energy was set to 538.1 eV (537.1 eV inside the liquid to account for the 1eV escape barrier) with a full width at half maximum of 1.74 eV. The beta values vary significantly over the width of the O1s band (increasing with eKE by 0.2 over the FWHM for hv ≈ 550 eV). As in the experiment the calculated $\beta_M$ for the O1s spectra represent averages over the band width. For the valence band spectra, by contrast, Fig. 2B in the main manuscript shows the calculated $\beta_M$ directly as it varies over the band system. For the initial angular distribution of photoelectrons in the liquid for ionization from the valence orbitals we used the experimental anisotropy parameters for the n=6 cluster from ref. [24].

**Comparison of average $\beta$ at high eKEs and in the subexcitation range:**

At higher electron kinetic energies (>> 10 eV), the average $\beta$ over a band in a photoelectron spectrum is roughly determined by the ratio of the CS for electronically inelastic scattering to the quasi-elastic momentum transfer CS [22]. In other words, this is the ratio of the probability that the electron is lost for detection to the probability for randomization of its direction for this particular band. For sub-excitation electrons (see below) where vibrationally inelastic scattering increasingly distorts the photoelectron band shape, the relation between the average $\beta$ over a band and the individual scattering channels becomes more complicated. The reason is the pronounced variability of the angular and energy loss characteristics of the individual channels in this energy range, which causes the above-mentioned pronounced variation of $\beta$ over a single photoelectron band.

## S3. Experiment and modelling of VMIs

**Experimental VMIs:**

Photoelectron velocity map images (VMIs) were recorded for water droplets after EUV excitation at more than 40 different photon energies below ~15.4eV [18,29] (e.g. Fig. 3A in the main manuscript). A three-plate VMI setup and a position sensitive detector were used for photoelectron detection. Details on the experimental setup and droplet VMIs can be found in refs. [18,33-35].

**Calculated VMIs:**

The model mimics the experimental VMI conditions and calculates directly the two-dimensional images on the position sensitive detector (e.g. Fig. 3A in the main manuscript). The light intensity inside the droplets is calculated from the complex index of refraction by solving Maxwell's equations numerically using a finite-difference time-domain (FDTD) code [FDTD solutions from Lumerical Solutions Inc. (www.lumerical.com).]. For details see refs. [18,29,34]. The Monte Carlo model for electron scattering described in section S4 and the recommended liquid CSs described in section S5 are used for the scattering calculations. The liquid CSs in section S5 correspond to refitted CSs from our previous work [18,29].

## S4. Electron scattering model

The scattering model is explained in detail in refs. [18,24,29] and the corresponding supplementary information, and is only briefly summarized here. The computer implementation of the model is based on a Monte-Carlo solution of the transport equation. A few million trajectories are sufficient for low resolution survey eKE spectra, the simulation of a detailed angle resolved spectrum requires up to a billion trajectories. This becomes possible with a highly parallel computer program which also allows the refinement of model parameters using a grid based Marquardt-Levenberg algorithm.

The probability to generate a conduction band electron at a certain location in the liquid jet or droplet is proportional to the local light intensity of the ionizing radiation at this location. The local light intensity is calculated from Maxwell's equations for plane wave irradiation. Further specific details for liquid micro-jets and droplets are mentioned above. The probabilistic electron transport model is formulated as a random walk with an exponential distribution of step lengths. The mean step length, i.e. the electron mean free path MFP($E$), is given by,

$$\text{MFP}(E) = \frac{1}{\rho \sigma_{\text{tot}}(E)}$$

where $E$ is the total energy of the electron, $\rho$ is the number density of scatterers (water molecules) and $\sigma_{\text{tot}}(E)$ is the total scattering cross section. The different scattering events are described by differential scattering cross sections (DCS), $\sigma(E, \Delta_E, \Omega)$, for the energy loss $\Delta_E$ and the deflection angle $\Omega$ of the electron. $\sigma(E, \Delta_E, \Omega)$ is written as the sum of contributions from elastic ($\Delta_E = 0$) and different types of inelastic scattering channels (inelastic electron-phonon, electron-vibron, dissociative electron attachment, electron-electron scattering):

$$\sigma(E, \Delta_E, \Omega) = \frac{1}{4\pi} \sum_i \sigma_i(E) g_i(\Delta_E) \{1 - \gamma_i(E)[1 - 2h(\cos\theta)]\}$$

$g_i(\Delta_E)$ is the energy loss distribution and $\gamma_i(E)$ the relative contribution of forward scattering in channel $i$. The $g_i(\Delta_E)$ take the form of Gaussian distributions with characteristic centers $\delta_i$ and widths $b_i$ (assumed to be independent of $E$):

$$g_i(\Delta_E) \propto exp\{-(\Delta_E - \delta_i)^2/b_i^2\}$$

$h(\cos\theta)$ is the Heaviside function. Note that in our previous work on electron scattering in water [18,29] an additional $\cos\theta$-dependence of the forward scattering contribution put higher weight on small deflection angles than in amorphous ice. The effect on predicted effective EALs and $\beta$ values is very small (no more than a few per cent). The above definition now follows exactly that for amorphous ice of refs. [25,26]. The total cross section is given by $\sigma_{\text{tot}}(E) = \iint \sigma(E, \Delta_E, \Omega) \, d\Delta_E d\Omega$. The construction of the DCSs for liquid water is described in the following section S5, building on DCSs determined for amorphous ice by Sanche and coworkers [25,26]. The DCSs are for an escape barrier of $V_0 = -1\text{eV}$ with respect to the vacuum level.

## S5. Electron scattering cross sections for liquid water

The liquid water cross sections are derived from the amorphous ice cross section of refs. [25,26] with extensions in the sub-excitation range and for electron energies > 100eV. Below 100 eV, the energy dependent functions $\sigma_i(E)$ and $\gamma_i(E)$ are parameterized in terms of their values at a discrete set of sampling points in $E$ and interpolation in between (double-logarithmic for $\sigma_i$ and linear for $\gamma_i$). As sampling points, we chose the tabulated values of ref. [25] augmented by those of ref. [26]. Ref. [26] only quotes the (isotropic) elastic and the total inelastic CS. To obtain CSs for individual channels, we assumed the relative contribution of individual inelastic channels as well as their $\delta_i$, $b_i$, and $\gamma_i(E)$ to be the same as in ref. [25] (with $\gamma_i(E)$ inter/extrapolated where necessary). In the region of overlap ($E < 20$eV), new parameters were obtained as averages of the two sets. Above 20 eV, the parameters remain unchanged from those of ref. [25]. Table S1 lists different CSs at the sampling points of the new augmented parameter set. We include the quasi-elastic momentum transfer CS, $\sigma_m$, since its ratio to the electronically inelastic CS largely determines the photoemission anisotropy beyond the sub-excitation range. $\sigma_m$ is given by

$$\sigma_m = \int (1 - \cos\theta) \, \sigma_{\text{quasielastic}}(\Omega) d\Omega = \sum_i \sigma_i(E)(1 - \tfrac{1}{2}\gamma_i)$$

Where the sum extends over all quasi-elastic channels (elastic, phonon, vibron). Note that $E$ refers to the electron's total energy with respect to the vacuum level, so that its kinetic energy in the liquid is given by $E - V_0$.

**Retrieval and extension of CS data:**

(1) Sub-excitation range ($E < 7$eV): The cross sections for liquid water were directly extracted from fits of calculated droplet VMIs to experimental VMIs. The CSs retrieved here are the result of refitting the CSs of our previous work [18,29]. The values at the sampling points served as fit parameters. We only refined the (isotropic) elastic and the total inelastic CS, keeping the relative contributions of individual inelastic channels (phonon, vibron, electronic) as well as the other parameters ($\delta_i, b_i, \gamma_i(E)$) at their amorphous ice values [25,26] (interpolated/extrapolated where necessary, see above). As before a single additional sampling point at $E = 0$ eV proved sufficient to describe the electron scattering at the lowest energies. We also confirmed our previous finding that the experimental droplet VMIs are already very well reproduced by the simulations without modifying the values at the other sampling points from those of amorphous ice. We also tried to vary the escape barrier, but did not find any improvement. Liquid water and amorphous ice CSs are apparently very similar also in the sub-excitation regime. Consistent with this observation, we fixed the liquid CS values at the sampling points to the corresponding ice values and keeping the value of $V_0$ (of $V_0 = -1$eV). See also the supporting information of ref. [18] for a further description of the retrieval method from VMIs.

(2) 7 eV to 100 eV: As explained in the main manuscript, experimental results currently available in this energy range agree well within experimental uncertainties with predictions on the basis of amorphous ice scattering cross sections. In the absence of more accurate experimental data no meaningful improvement would result from a fit. Consequently, we keep the parameter values (i.e. $\delta_i$ and $b_i$ as well as $\sigma_i(E)$ and $\gamma_i(E)$ at the sampling points) frozen at their amorphous ice values [25,26].

(3) Above 100 eV: The differential cross section for electron loss ($\sigma_{other}(E)$ in ref. [25]) subsumes all electronically inelastic scattering channels (electronic excitation, impact ionization etc.). To extrapolate to higher electron energies, we make use of the energy dependence of the (electronically) inelastic MFP $\lambda_{FPA}$ predicted by Shinotsuka et al. using their relativistic full Penn algorithm [10]. Requiring the CS at 100 eV to match that of Michaud et al. [25] leads to

$$\sigma_{other}(E) = \sigma_{other}(100\text{eV}) \cdot \frac{\lambda_{FPA}(100\text{eV})}{\lambda_{FPA}(E)}$$

for $E > 100\text{eV}$. To obtain values for the quasi-elastic CS (elastic, phonon, vibron), we assume that the ratio of the quasi-elastic momentum transfer CS, $\sigma_m$, to the electronically inelastic CS scales with $E$ in the liquid in the same way as in the gas phase. We exploit that above 100 eV the electronically inelastic channel in the gas phase is dominated by the total ionization CS, $\sigma_{ion}^g$, which we take from table 11 of ref. [36]. Further assuming the relative contributions of the various quasi-elastic channels to remain constant above 100 eV leads to the following extrapolation of individual quasi-elastic CS

$$\sigma_i(E) = \sigma_i(100\text{eV}) \cdot \frac{\lambda_{FPA}(100\text{eV})}{\lambda_{FPA}(E)} \cdot \frac{\sigma_m^g(E) \cdot \sigma_{ion}^g(100\text{eV})}{\sigma_m^g(100\text{eV}) \cdot \sigma_{ion}^g(E)}$$

At the relatively high energies (and the correspondingly small de Broglie wavelengths), this should be a reasonable approximation. The gas phase momentum transfer CS, $\sigma_{ion}^g$, is extrapolated double-logarithmically from table 5 of ref. [36]. Furthermore, we assume the remaining parameters ($\delta_i, b_i, \gamma_i$) for the various scattering channels to remain constant above 100eV.

**Estimated uncertainties of liquid water CS:**

(1) Sub-excitation range ($E < 7\text{eV}$): As in our previous work [18,29] we estimate an overall uncertainty of about a factor of two, i.e. -50%/+100%, for the absolute values of the isotropic elastic and total inelastic CSs for electron scattering in the liquid. The lower bound is comparable to the error quoted for the absolute value of the amorphous ice CS at electron energies below 20 eV [25]. We quote slightly more generous error bounds for the liquid to account for additional uncertainties in the shape of the experimental VMIs from which the liquid values were retrieved in that energy range.

(2) 7 eV-100 eV: In this range, we recommend to adopt the amorphous ice data from refs. [25,26] unchanged. For these, Michaud et al. [25,26] quote a relative uncertainty of raw data (relative values as a function of electron energy) of 5% for eKE < 20 eV, increasing progressively up to 20% at eKE = 100 eV. For the additional uncertainty of the CS scale (absolute values) they estimate a systematic error of 25%. For the total CS this yields an uncertainty of 30-45%. On this basis, we chose the error bars of our simulations. For the « effective EALs » this corresponds to an uncertainty of the total CS of 45%, while for $\beta$ values this corresponds to an uncertainty of 20% each for the quasielastic CS and for the electronically inelastic (=loss) CS (i.e. roughly 40% for the ratio of the latter two). Evidently, we cannot quantify the error arising from any deviation between ice and liquid CS. Judging, however, from the good agreement of the predicted « effective EALs » and $\beta$ values with measured values well within experimental uncertainties, this does not seem to be a significant source of error.

(3) above 100 eV: It is not possible to quantify the uncertainty of the proposed extrapolation of CS beyond 100 eV. Again, judging from the good agreement of predicted beta values and effective EALs for eKE > 100 eV with experimental results (and keeping in mind that no fitting is involved) the proposed extrapolation does not seem to be unreasonable. For the error bars of our simulations, we have therefore used the same error estimates as in the range 7eV $\leq$ eKE $\leq$ 100eV. Predicted effective EALs (or rather effective photoemission yields) come to lie at the lower end of experimental uncertainties, which - if anything - might suggest that the extrapolated CSs tend to be slightly too high.

**Can ad hoc scaling improve the description of electron scattering in the liquid?**

It has repeatedly been suggested that currently available ice cross sections from refs. [25,26] should be scaled – invariably by factors significantly larger than unity - to obtain CSs that would be applicable to electron scattering in the liquid (see e.g. discussion in refs. [5,6,12,27,28]). In some cases, a suggested very large up-scaling of the elastic CS seems to have been the result of confusing the total elastic CS with the isotropic elastic CS quoted for amorphous ice (see e.g. discussion in refs. [5,6] and references therein). Elastic scattering is dominated by a sharp forward peak, which does not contribute to electron transport. Instead, the effective contribution of elastic scattering to electron transport is exactly equivalent to isotropic scattering with the (elastic) momentum transfer cross section (see above). The isotropic elastic CS quoted for amorphous ice [25,26] thus represents a momentum transfer cross section. The elastic CS in the liquid is unknown. In the gas phase the difference between the (rotationally) elastic and the corresponding momentum transfer CS amounts to a factor of 3-4 between for 10 < $E$ <100 eV, but easily exceeds an order of magnitude in the sub-excitation range [36]. There does not seem to be any a priori justification for applying ad hoc scaling factors to the ice CSs. On the basis of the comparison with currently available experimental results for the liquid (effective EAL, Fig.1, and $\beta$-parameters, Fig.2), scaling factors beyond two can be excluded a posteriori. This leaves the question open, whether smaller scaling factor might provide a meaningful improvement over just using amorphous ice CS.

Sub-excitation range ($E$ < 7eV): There is no rationale for upscaling amorphous ice CSs in the sub-excitation range. It would not improve the agreement with droplet VMIs (Fig. 3). Unscaled amorphous ice CSs have even been reported to result in a significantly better agreement with experimental thermalization lengths in the liquid compared with twofold increased CSs [12].

Up to a few 10 eV: In the region around a few 10 eV, models based on dielectric response functions have been cited to suggest both upscaling and downscaling of ice cross sections, depending on the model [10,11,13-16]. The applicability of such models in this eKE range is, however, disputable. The comparison with experimental results in Figs. 1 and 2 does not provide any hint for a systematic under- or overestimation of CSs in the liquid. The deviations between ice predictions and experimental liquid data lie well within experimental uncertainties. If a meaningful trend could be extracted at all, Fig. 1 might suggest that the total CS in the liquid for eKE above ~30 eV lies slightly below that of amorphous ice.

Above a few 10 eV: Models for the liquid based on dielectric response functions generally predict electronic IMFP's below the corresponding ice IMFPs in the eKE range above a few 10 eV [10,11,13], suggesting that upscaling of the corresponding ice CSs may be required. However, the comparison of predicted and experimental EALs in Fig. 1 seems to point in the opposite direction. If at all, these data would seem to suggest downscaling rather than upscaling in the region above ~30 eV.

From the above considerations it must be concluded that no meaningful improvement of the description of electron scattering in the liquid can be expected from any ad hoc scaling over simply using the unchanged amorphous ice cross sections. At the current state of knowledge the scattering behavior of electrons in the liquid appears to be indistinguishable from that in amorphous ice.

**Table S1: CSs for electron scattering in liquid water.** $\sigma_{tot}$ = total scattering CS, $\sigma_{inel}$ = total inelastic scattering CS, $\sigma_m$ = quasi-elastic momentum transfer CS. $E$ refers to the electron's total energy with respect to the vacuum level as in the refs. [25,26]. The data is for an escape barrier of $V_0 = -1$eV. The electron kinetic energy in the liquid is given by $E - V_0$. The corresponding mean free paths are $\text{MFP}(E) = \frac{1}{\rho \sigma_i(E)}$ where $\rho = 3.343 \cdot 10^{22} \text{cm}^{-3}$ is the number density of the water molecules (see Fig. S2).

| $E$ / eV | $\sigma_{tot}$ / $10^{-16}$cm$^2$ | $\sigma_{inel}$ / $10^{-16}$cm$^2$ | $\sigma_m$ / $10^{-16}$cm$^2$ |
|---|---|---|---|
| 0.0 | 2.252 | 1.652 | 1.787 |
| 1.2 | 1.371 | 1.077 | 1.068 |
| 1.7 | 1.103 | 0.844 | 0.863 |
| 2.2 | 1.007 | 0.770 | 0.786 |
| 3.2 | 0.815 | 0.609 | 0.639 |
| 4.2 | 0.922 | 0.646 | 0.738 |
| 5.2 | 0.998 | 0.681 | 0.796 |
| 6.2 | 1.072 | 0.753 | 0.844 |
| 7.2 | 1.261 | 0.914 | 0.964 |
| 8.2 | 1.158 | 0.873 | 0.862 |
| 9.2 | 1.030 | 0.798 | 0.714 |
| 10.2 | 0.865 | 0.638 | 0.604 |
| 11.2 | 0.810 | 0.558 | 0.573 |
| 12.2 | 0.911 | 0.613 | 0.633 |
| 13.2 | 1.056 | 0.699 | 0.723 |
| 14.2 | 1.174 | 0.782 | 0.774 |
| 15.2 | 1.230 | 0.843 | 0.769 |
| 16.2 | 1.147 | 0.839 | 0.662 |
| 17.2 | 1.199 | 0.937 | 0.628 |
| 18.2 | 1.062 | 0.859 | 0.523 |
| 19.2 | 0.999 | 0.842 | 0.458 |
| 20.0 | 1.056 | 0.855 | 0.497 |
| 22.5 | 1.080 | 0.914 | 0.452 |
| 25.0 | 1.090 | 0.949 | 0.413 |
| 27.5 | 1.118 | 0.984 | 0.403 |
| 30.0 | 1.141 | 1.009 | 0.397 |
| 32.5 | 1.222 | 1.085 | 0.412 |
| 35.0 | 1.264 | 1.126 | 0.416 |
| 37.5 | 1.300 | 1.162 | 0.418 |
| 40.0 | 1.296 | 1.163 | 0.403 |
| 42.5 | 1.305 | 1.175 | 0.394 |
| 45.0 | 1.305 | 1.180 | 0.379 |
| 47.5 | 1.284 | 1.165 | 0.363 |
| 50.0 | 1.221 | 1.113 | 0.334 |
| 52.5 | 1.278 | 1.171 | 0.340 |
| 55.0 | 1.288 | 1.186 | 0.333 |
| 57.5 | 1.262 | 1.168 | 0.319 |
| 60.0 | 1.276 | 1.189 | 0.314 |
| 62.5 | 1.312 | 1.231 | 0.315 |
| 65.0 | 1.319 | 1.245 | 0.308 |
| 67.5 | 1.286 | 1.221 | 0.291 |
| 70.0 | 1.424 | 1.360 | 0.310 |
| 72.5 | 1.528 | 1.466 | 0.321 |
| 75.0 | 1.550 | 1.492 | 0.313 |
| 77.5 | 1.578 | 1.524 | 0.307 |
| 80.0 | 1.673 | 1.622 | 0.312 |
| 85.0 | 1.650 | 1.604 | 0.300 |
| 90.0 | 1.518 | 1.478 | 0.270 |
| 100.0 | 1.496 | 1.459 | 0.261 |
| 200.3 | 0.979 | 0.964 | 0.101 |
| 298.9 | 0.743 | 0.734 | 0.061 |
| 403.4 | 0.595 | 0.589 | 0.042 |
| 492.7 | 0.512 | 0.507 | 0.034 |
| 601.8 | 0.439 | 0.435 | 0.026 |
| 735.1 | 0.375 | 0.372 | 0.021 |
| 1096.6 | 0.272 | 0.270 | 0.013 |

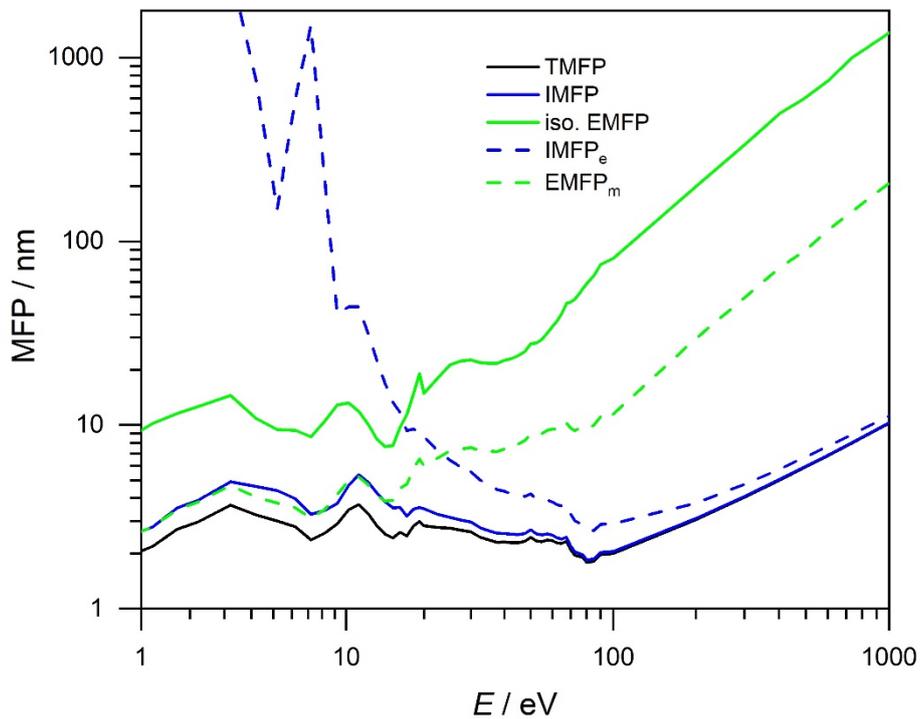

**Fig. S2. Different mean free paths (MFPs) for liquid water.** TMFP = total MFP, IMFP = total inelastic MFP, $IMFP_e$ = electronically inelastic MFP, iso. EMFP = isotropic elastic MFP, $EMFP_m$ = quasi-elastic momentum transfer MFP.